\documentclass[preprint,12pt]{elsarticle}

\journal{Physica A}

\usepackage{graphicx}
\usepackage{amsmath, amssymb}
\usepackage{lmodern}
\usepackage[T1]{fontenc}
\usepackage[utf8]{inputenc}
\usepackage[english]{babel}
\usepackage{color}

\begin{document}

\begin{frontmatter}
\title{Influence of the neighborhood on cyclic models of biodiversity}

\author[PAR]{D. Bazeia}
\author[PAR]{M. Bongestab}
\author[MAR]{B.F. de Oliveira}

\address[PAR]{Departamento de F\'\i sica, Universidade Federal da Para\'\i ba, 58051-970 Jo\~ao Pessoa, PB, Brazil}
\address[MAR]{Departamento de F\'\i sica, Universidade Estadual de Maring\'a, 87020-900 Maring\'a, PR, Brazil}

\begin{abstract}
This work deals with the influence of the neighborhood in simple rock-paper-scissors models of biodiversity. We consider the case of three distinct species which evolve under the standard rules of mobility, reproduction and competition. The rule of competition follows the guidance of the rock-paper-scissors game, with the prey being annihilated, leaving an empty site in accordance with the May-Leonard proposal for the predator and prey competition. We use the von Neumann neighborhood, but we consider mobility under the presence of the nearest, next nearest and next to next nearest neighbors in three distinct environments, one with equal probability and the others with probability following power law and exponential profiles. The results are different, but they all show that increasing the neighbourhood increases the characteristic length of the system in an important way. We have studied other models, in particular the case where one modifies the manner a specific species competes, breaking the cyclic evolution and unveiling the interesting result in which the strongest individuals may constitute the less abundant population.
\end{abstract}


\end{frontmatter}

\section{INTRODUCTION}

Biodiversity is a very complex subject, and biodiversity maintenance is one of the most challenging issues in science in general. In spite of this, interesting specific works developed in the last three decades have shown how a very simple model may contain important information to sustain biodiversity \cite{1996-Sinervo-Nature-380-240, 2002-Kerr-N-418-171, 2004-Kirkup-Nature-428-412, 2007-Reichenbach-N-488-1046, 2014-Weber-JRSI-11-96, 2020-Liao-N-11-6055}. An interesting result is that a three-species model that evolves under the guidance of mobility, reproduction and competition, with the cyclic rules of competition following the rock-paper-scissors game is capable of offering the basic ingredients to control biodiversity in nature \cite{1996-Sinervo-Nature-380-240,2002-Kerr-N-418-171,2004-Kirkup-Nature-428-412}.

These and other related works have stimulated several investigation in the subject, as we see from the recent reviews in the area \cite{2007-Szabo-PR-446-97, 2014-Szolnoki-JRSI-11-0735, 2018-Dobramysl-JPA-51-063001, 2020-Szolnoki-EPL-131-68001}. One of the simplest systems consists of the three distinct species $(A, \textrm{red})$, $(B, \textrm{blue})$ and $(C, \textrm{yellow})$, which evolve under the stochastic rules of mobility $(m)$, reproduction $(r)$ and competition $(p)$, with $m+r+p=1$. The time evolution in general occurs in a square lattice of $N\times N$ sites, which obey periodic boundary conditions, with each site being surrounded by a neighborhood of the four nearest sites, two at the left and right $(i\pm1,j)$ and two at the top and bottom $(i,j\pm1)$ of the given site $(i,j)$. This is known as the von Neumann neighborhood, and is frequently used to implement the numerical simulations described in the related literature, as we can find, for instance, in Refs. \cite{2008-Peltomaki-PRE-78-031906, 2008-Reichenbach-JTB-254-368, 2010-He-PRE-82-051909, 2010-Yang-C-20-2, 2011-He-EPJB-82-97, 2012-Avelino-PRE-86-031119, 2012-Avelino-PRE-86-036112, 2012-Roman-JSMTE-2012-p07014, 2013-Roman-PRE-87-032148, 2013-Vukov-PRE-88-022123, 2014-Avelino-PRE-89-042710, 2014-Szolnoki-PRE-89-062125, 2017-Brown-PRE-96-012147, 2017-Park-SR-7-7465, 2018-Shadisadt-PRE-98-062105} and in references therein.

Among the rules of mobility, reproduction and competition, the last one plays an important role to control biodiversity, requiring a non hierarchical environment where it is controlled by the cyclic rules of the rock-paper-scissors game, in which scissors cuts paper, paper wraps rock and rock breaks scissors. Furthermore, in Ref. \cite{2007-Reichenbach-N-488-1046} the authors revealed that mobility also plays an important role, since it may be used to break biodiversity. This happens when one increases the mobility parameter $m$ toward unity, consequently increasing the characteristic length of the system to higher and higher values. This behavior was studied before in several distinct scenarios, always confirming the fact that mobility defies biodiversity; see, e.g., Refs. \cite{2011-Jiang-PRE-84-021912, 2013-Park-Chaos-23-023128, 2014-Cheng-SR-4-7486, 2017-Bazeia-EPL-119-58003, 2018-Avelino-PRE-97-032415} and references therein.

\begin{figure}[ht]
	\centering
		\includegraphics[width=4.2cm]{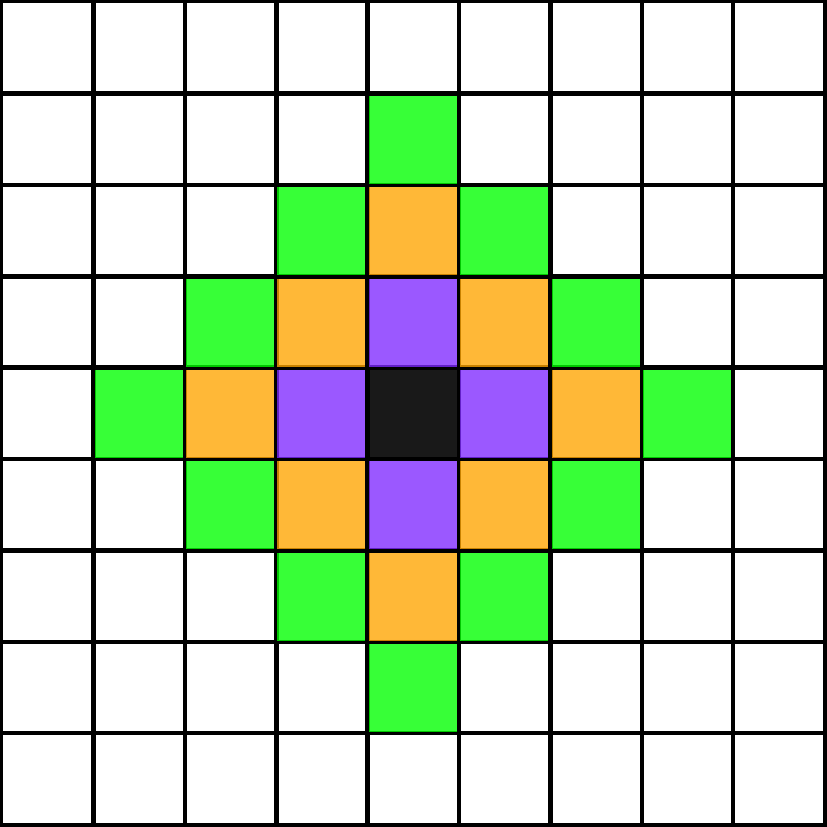}
	\caption{Illustration of the first, second and third neighbourhood with the nearest (violet), next nearest (orange) and next to next nearest (green) neighbours, respectively, in the von Neumann neighborhood which we consider in this work.}
	\label{fig1}
\end{figure}

Mobility has also been studied in several other contexts: for
instance, in \cite{1995-Vicsek-PRL-75-1226,2016-Cambui-PA-4444-582}
the authors have investigated the behavior of systems of
self-propelled particles, in which the constituent particles are
driven to move in the average direction of the particles in their
neighborhood. Also, in Ref. \cite{2018-Dobramysl-JPA-51-063001} the
investigation describes a diversity of aspects of spatially extended
predator–prey systems, and in \cite{2018-Avelino-PRE-97-032415} the
study deals with directional mobility, in which the rule of mobility
is modified to let the species move following the direction associated
to a larger number of target preys in the surrounding neighborhood. In
this case, the motivation was to add a kind of eye or nose for the
predator to see or smell the prey, contributing to make the system
more realistic when one thinks to describe biodiversity in nature in
several different contexts. Despite the important achievements already
developed with three species and the simple mobility, reproduction and
competition rules, we know that the competition and reproduction rules
in the wild may be very diverse and sometimes hard to match with the
rules used in lattice simulations. A typical situation occurs, for
instance, in the outcome of competition among plant species, due to an
effect of dispersal on the spatial distribution of individuals
\cite{Pacala1986}, and also in the hunting of fishes, since the prey
aggregations in fish schools may sometimes attract several predators,
and they hardly behave as the one-to-one active and passive neighbors,
which is in general used in numerical simulations; see, e.g.,
\cite{Gtmark1986,Thiebault2015} and references therein for several
studies on the capture of fishes in a fish school. {\color{black}We also add the interesting study on the asymmetric interactions among strains of bacteria \cite{2020-Liao-N-11-6055} which induces the survival of the weakest \cite{2001-Frean-PRSLB-268-1323}, and another system of bacteria communities which investigates the competition between antibiotic producing bacteria, non-producers (or cheaters), and sensitive cells \cite{pone}. In the last case, the authors described the presence of a range of growth rates of the resistant cells where the species coexist, and the production mechanism appears to be maintained as a polymorphism in the producing species, with the resistance mechanism maintained as a polymorphism in the sensitive species.} These results have motivated us to contribute to describe more realistic models, and in this work we explore other possibilities, in which the neighborhood is generalized to accommodate interactions with the nearest, the next nearest and the next to next nearest neighbors. This is illustrated in Fig. \ref{fig1} and there we count four nearest, eight next nearest and twelve next to next nearest neighbors.

In order to investigate how the neighborhood modifies the time evolution of the systems under consideration, we organize the work as follows: in the next Sec. \ref{model} we describe the rules to be used to investigate the models that account for the distinct neighborhood with distinct probabilities. We then proceed to Sec. \ref{result}, where the results obtained in the numerical simulations are displayed and discussed; moreover, in Sec. \ref{other} we investigate other possible usages of the extended neighborhood in order to break the behavioral symmetry among the species of the model as we discuss how this contributes to jeopardize biodiversity. We conclude the work in Sec. \ref{conclusion}, including some new lines of investigations related to the main results of the present work

\section{The models}
\label{model}

In this work we will investigate a model with three distinct species, marked by the colors red (species $A$), blue (species $B$) and yellow (species $C$), which evolve in time according to stochastic rules of mobility $(m)$, reproduction $(r)$ and competition $(p)$.  We will in fact consider the simpler case where $r=p=(1-m)/2$; thus, one just needs to give $m$ to define the three rules of mobility, reproduction and competition. Here, we have used $m=0.5$ unless otherwise specified. The model evolves in a square lattice of linear size $N$ containing $N \times N$ sites that obey periodic boundary conditions. We have used $N= 1000$, unless otherwise specified. In the lattice, we create an initial state by randomly distributing the three species and the empty sites, identified by the color white. The rules used to describe the time evolution of the model are the standard ones: first, a site is randomly selected, which is named the active site. If the active site is empty, we return and choose another site; if the active site is one of the three species, we randomly choose a rule $m$, $r$ or $p$ and then the passive site. If mobility is chosen, we exchange the active and passive sites, but if reproduction is chosen, and if the passive site is empty, we then color it with the color of the active site. The rules of competition are different, and they follow the rock-paper-scissors game. Here $A$  outcompetes $B$, $B$ outcompetes $C$ and $C$ outcompetes $A$ in the cyclic and non hierarchical way, as required. In this sense, if the active site is $A$  and the passive $B$, the passive site becomes white, that is, the $B$  individual is annihilated, leaving the corresponding site empty. This follows the instruction of the May-Leonard predator and prey rule. The procedure is similar for $B$ and $C$, and for $C$ and $A$, respectively. The time evolution of the simulations to be described in this work is counted in the form of generations, with a generation being the time necessary for the system to undergo $N^2$ specific accesses to the lattice. 

The basic model to be considered in this work engenders similar rules of reproduction and competition, at the range of the four nearest neighbours. However, we will consider five distinct scenarios for the rule mobility. The first three of them falls in the same family: model $a_1$ is the standard model, with the mobility occurring only with the four nearest neighbors; model $a_2$, with the mobility occurring with the nearest and next nearest neighbours; model $a_3$, with the mobility occurring with the nearest, next nearest and next to next nearest neighbours. See Fig. \ref{fig1} for an illustration of the three neighbourhoods. In the model $a_2$, when mobility is selected one then picks the first or the second neighborhood with probability $1/2$, and in model $a_3$ the probability is $1/3$, to select the first, second or third neighborhood. We have considered another family of models, where the interaction with the second and third neighbor follow the $1/r$ power law behavior. We then added two other models with the $1/r$ profile, namely, models $b_2$ and $b_3$. In the first one, the $b_2$ model, the probabilities to interact with the nearest and next nearest individuals are $2/3$ and $1/3$, respectively. And in model $b_3$ the corresponding probabilities are controlled by $0.55$, $0.28$ and $ 0.17$, respectively. These probabilities have to add to unity and follow the $1/r$ profile, with $r=1,2,3$ identifying the nearest, next nearest, and next to next nearest neighbourhood, respectively. Evidently, inside the same neighborhood, the passive site is chosen with the very same probability.

As we have already informed, the motivation to study these four new models is to capture their behavior when the species are allowed to evolve under the possibility of interacting with the other species in a qualitatively different manner, where its range may be somehow farther away from being the first neighbour. Since in the five models to be considered in this work, mobility occurs with a much larger probability, when compared with reproduction and competition, we just considered mobility to see the first, second or third neighbor, since this appears to induce the main effect in the rules to be used in this work. 

\begin{figure}[!h]
	\centering
		\includegraphics{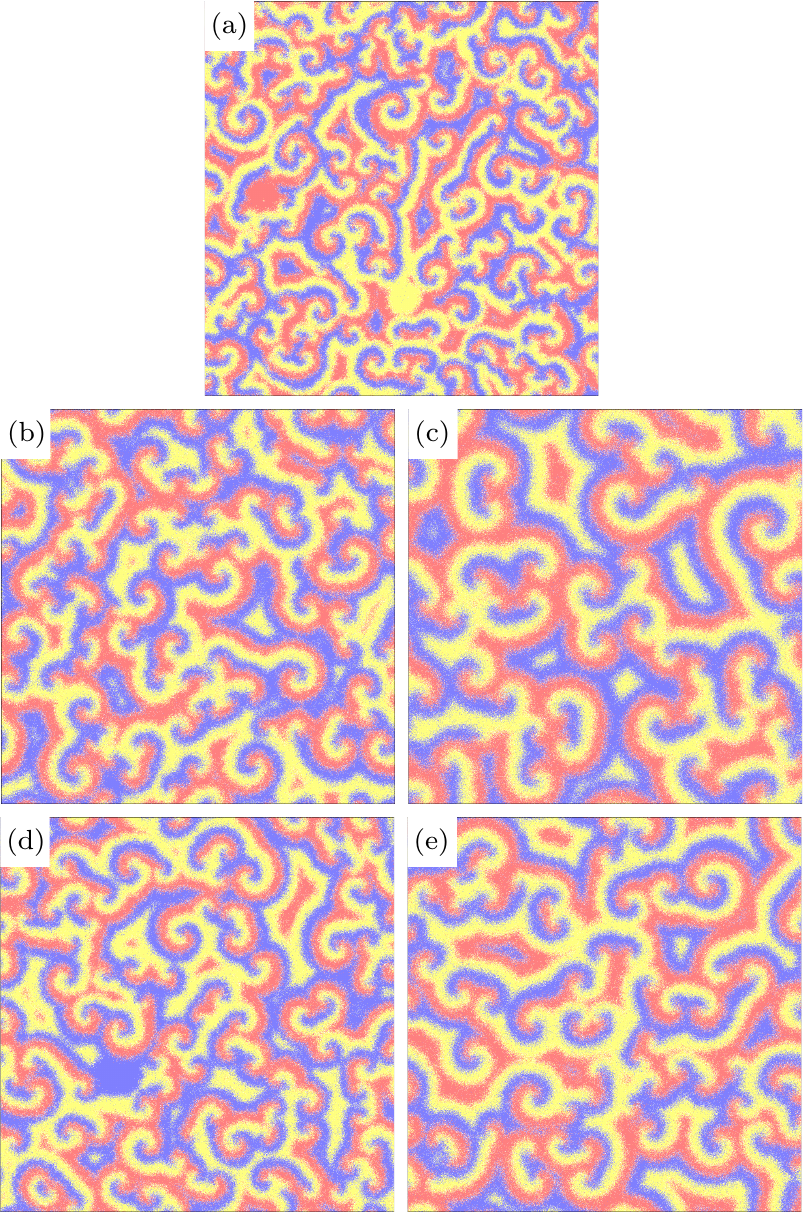}
	\caption{Snapshots of the configurations of the species $A$ (red), $B$ (blue), and $C$ (yellow) after running the numerical simulations for $10^4$ generations for the models $a_1$, $a_2$, $a_3$, $b_2$ and $b_3$ with $m=0.5$ and $N=1000$. Note the increase of the characteristic length of the system as we increase the neighborhood for the mobility of the individuals.}
	\label{fig2}
\end{figure}

\section{Results}
\label{result}

The first simulations implemented in this work concern the time evolution of the five models until
the generation $10^4$. The corresponding snapshots are depicted in Fig. \ref{fig2}, and there one notices that the characteristic length of the model is modified by the inclusion of other neighbors, and also by the modification of the probability associated to the presence of other next and next to next nearest neighbors.     

In order to quantify the spatial behavior which we identify in Fig. \ref{fig2}, we first define the scalar field $\phi(\vec{r}^{\;\prime})$ that represents  species in the position $\vec{r}^{\;\prime}$ in the lattice and we use the values $0$ for an empty site and $1$, $2$ and $3$ for the three species $A$, $B$ and $C$, respectively. We have used this before, for instance, in Refs. \cite{2018-Avelino-EPL-121-48003,2019-Bazeia-PRE-99-052408} and it is also possible to calculate $C(\vec r)$, which is the spatial autocorrelation function. It is given by the relation
\begin{equation}
	C(\vec{r}) = \dfrac{\mathcal{F}^{-1}\{\varphi(\vec{k})\varphi^*(\vec{k})\}}{C(0)} \ , 
	\label{eq1}
\end{equation}
where $\varphi(\vec{k}) = \mathcal{F}\{\phi(\vec{r}^{\;\prime}) - \langle \phi \rangle\}$, $\mathcal{F}$ being the Fourier transform and $\langle \phi \rangle$ standing for the mean value of the scalar field. We go further and instead of $C(\vec{r})$ we consider $C(r)$, using the following relation
 \begin{equation}
 	C(r) = \displaystyle \sum_{|r| = x+y} \dfrac{C(\vec{r})}{{\rm min}(2N - (x+y+1), x+y+1)} \ .
 	\label{eq2}
 \end{equation}

\begin{figure}[t]
	\centering
		\includegraphics{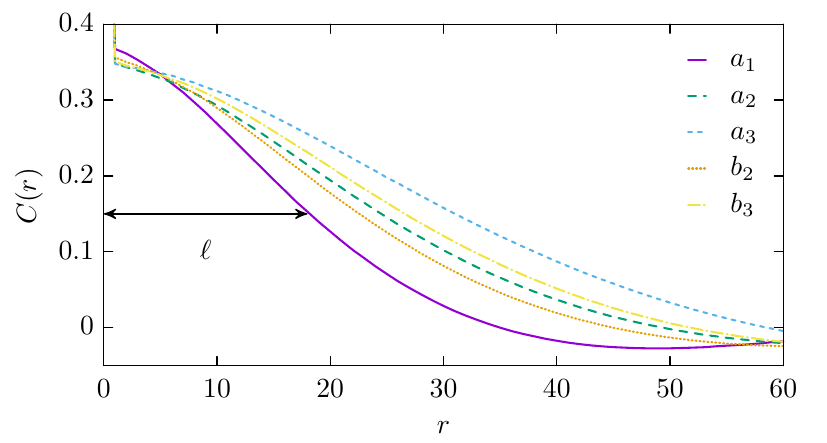}
	\caption{The spatial autocorrelation function for the five models considered in the work. The characteristic length is defined as $\ell$, with $C(r=\ell) = 0.15$, and it increases as one increases the neighborhood interaction. Here we have used $m=0.5$ and $N=1000$}.
	\label{fig3}
\end{figure}

In Fig.~\ref{fig3} we depict the function $C(r)$ for the five distinct models that we are investigating in this work. The arrow in Fig. \ref{fig3} indicates the characteristic length scale, $\ell$, and in the present work we define by the expression $C(r=\ell) = 0.15$.

\begin{figure}[!htb]
	\centering
		\includegraphics{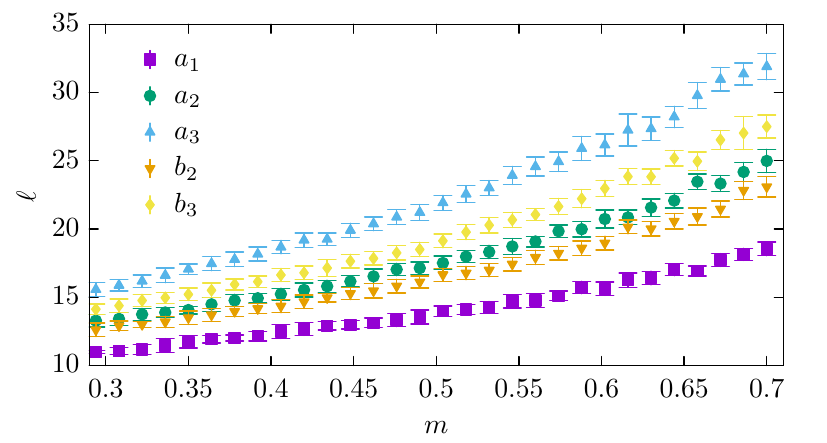}
	\caption{The characteristic length is shown as a function of mobility, for the five models. It increases in a nonlinear way, as we increase the mobility. Here we have used $m=0.5$ and $N=1000$.}
	\label{fig4}
\end{figure}

We now study the behavior of this characteristic length as a function of the mobility for the different models considered in this work. The results are displayed in Fig.~\ref{fig4}, and they show that if we increase the mobility the characteristic length also increase in a nonlinear way. As we can see, the increase in the number of neighbors makes the characteristic length also increase, and this follows the sequence: model $a_1$, model $b_2$, model $a_2$, model $b_3$, and model $a_3$, which leads us to the result that the larger the neighbourhood, the greater the characteristic length.    

\begin{figure}[!htb]
	\centering
		\includegraphics{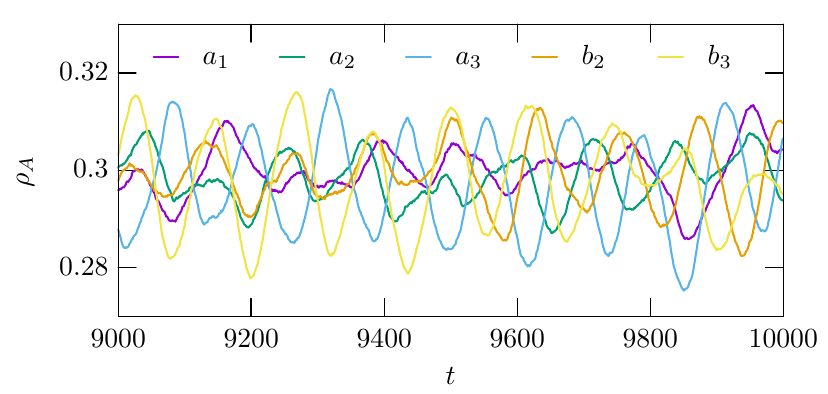}
	\caption{The abundance or density of species $A$ as a function of time is shown for the five models. Here we have used $m=0.5$ and $N=1000$. Note that, the noisiness of the density diminishes as we increase the neighborhood interaction.}
	\label{fig5}
\end{figure}

In order to have a more accurate view of the time evolution of the three species in the five different models, in Fig.~\ref{fig5} one depicts the abundance or density of one of the three  species, which we here choose to be the species $A$, as a function of time. A similar behavior is also found for the other species $B$ and $C$. In all models the abundance or density of the three species oscillate, so we also work to quantify such oscillations using the discrete Fourier transform, which is defined by 

\begin{equation}
    \rho(f)=\dfrac{1}{N_G} \sum_{t=0}^{N_G-1} \rho(t) \exp{(-2\pi i f t)}
 \end{equation}
where $f$ is the frequency and $N_G$ the number of generations.  In all cases, we have used 
$N_G= 10^4$ generations, but we have run the simulations for $15 \cdot 10^3$ generations and discarded the first $5 \cdot 10^3$ to ensure that the spiral pattern was already achieved.

\begin{figure}[!htb]
	\centering
		\includegraphics{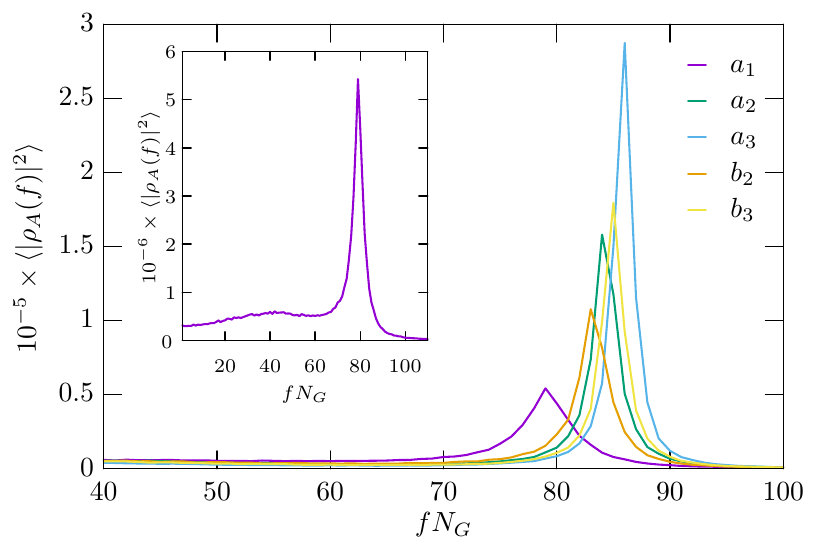}
	\caption{The power spectrum for species $A$. The position of the peak indicates how many oscillations there are in $10^4$ generations. Similar behaviour is also found for the other two species $B$ and $C$. Here we have used $m=0.5$ and $N=1000$}.
	\label{fig6}
\end{figure}

The results for the discrete Fourier transform are displayed in Fig.~\ref{fig6} for the five models. We are presenting the power spectrum averaged over $10^3$ simulations. The characteristic frequencies are $f_c = 79/N_G$ for model $a_1$, $f_c = 84/N_G$ for model $a_2$, $f_c = 86/N_G$ for model $a_3$, $f_c = 83/N_G$ for model $b_2$ and $f_c = 85/N_G$ for model $b_3$. The characteristic frequency is defined as the maximum value of the power spectrum, and its meaning is to indicate the approximate number of oscillations of the abundance or density of the corresponding species in the run of $10^4$ generations. One notices that the characteristic frequency depends on the neighborhood, and it increases as the neighborhood also increases. 

\begin{figure}[!htb]
	\centering
		\includegraphics{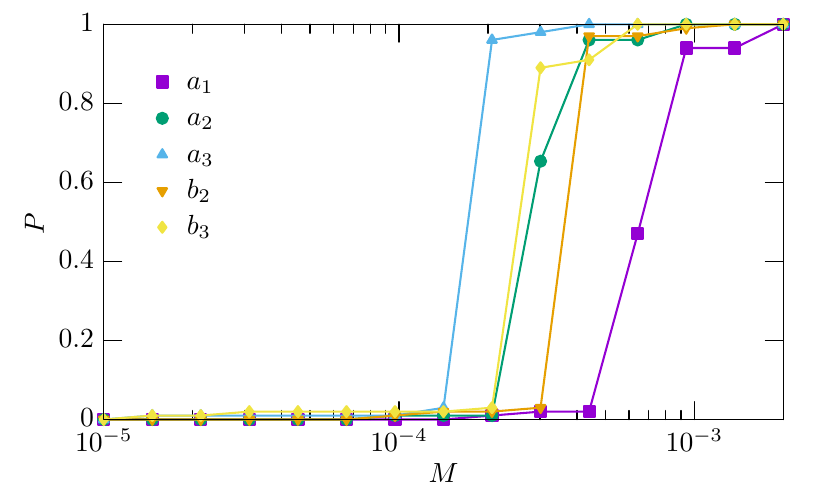}
	\caption{The extinction probability as a function of the mobility $M$. The extinction occurs earlier when we increase the neighborhood interaction. Here we have used a lattice with linear size $300$ and $10^3$ different initial conditions.}
	\label{fig7}
\end{figure}

We have also investigated the extinction probability, which is the probability of the model to break biodiversity. Since these simulations require much longer times, we have chosen to consider the linear size of the lattice as $N=300$; below we explain this choice. The results are in Fig. \ref{fig7}, and they show that the larger the neighborhood, the lower the value of the mobility $M$. This means that the increase of the neighbour contributes to increase the importance of mobility concerning breaking biodiversity. To implement this investigation, we changed the mobility $m$ to the new parameter $M$, which makes the mobility naturally proportional to the typical area explored by an individual per unit time. This follows as in \cite{2007-Reichenbach-N-488-1046}, and we considered the prescriptions described in the more recent works  \cite{2017-Bazeia-EPL-119-58003,2018-Avelino-PRE-97-032415}. The procedure requires that we change $p$, $r$, and $m$ to $p'$, $r'$, and $m'$, and rewrite the probabilities of competition, reproduction and mobility as $p'/(m'+p'+r')$, $r'/(m'+p'+r')$ and $m'/(m'+p'+r')$, respectively. Now, by setting $p'= r'= 1$, we can introduce the mobility parameter $M$ as $M =m'/2N^2$, which is proportional to the typical area explored by an individual.

\begin{figure}[!htb]
	\centering
		\includegraphics{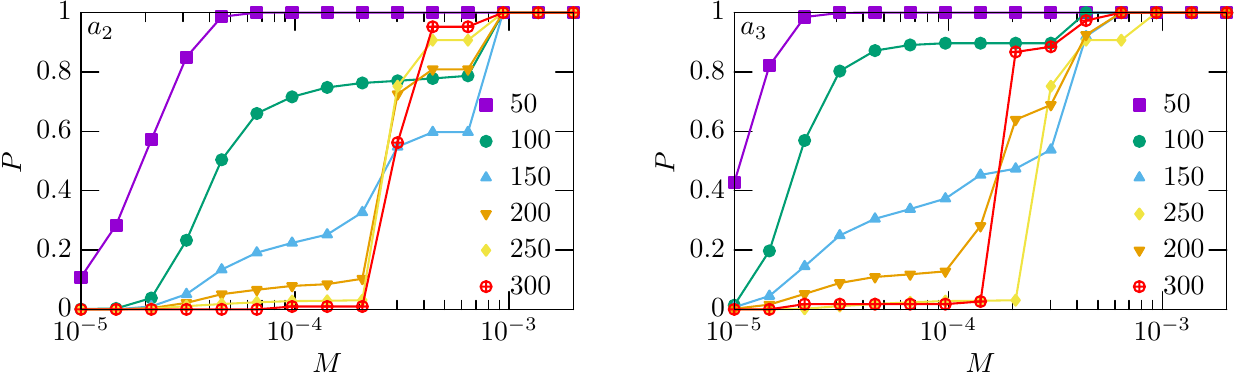}
	\caption{The extinction probability as a function of the mobility $M$ for the model $a_2$ (left) and $a_3$ (right). We have used lattices with linear sizes $50, 100, 150, 200$, $250$ and $300$, and $10^3$ different initial conditions.}
	\label{figX}
\end{figure}

Since we are using neighbourhood of distinct sizes, with nearest, next nearest and next to next nearest neighbours, it is important to show that the linear size $N=300$ is appropriate to avoid finite size effects in the simulations implemented in Fig. \ref{fig7}. Due to this, in Fig. \ref{figX} we display results for the extinction probability of the two models $a_2$ and $a_3$, which consider nearest and next nearest, and nearest, next nearest and next to next nearest  neighbours, respectively. And also, we consider lattices with distinct sizes, in particular the cases with linear size $N=50, 100, 150, 200$, $250$ and $N=300$. The numerical investigations show that the behavior of $M$ is qualitatively similar, and that the lattice size $N=300$ gives reliable results.

\begin{figure}[!htb]
	\centering
		\includegraphics{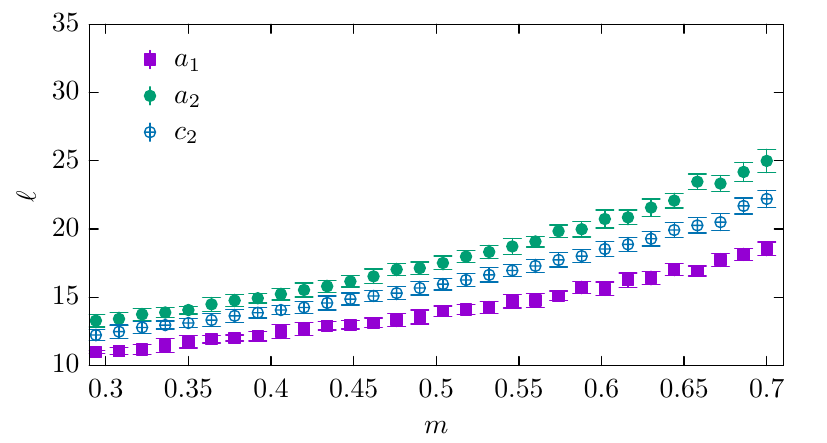}
	\caption{The characteristic length is shown as a function of mobility,  for the three models $a_1$, $a_2$, and $c_2$, for comparison. As in Fig. \ref{fig4}, here we also used $m=0.5$ and $N=1000$}.
	\label{fig8}
\end{figure}

The above investigations motivate us to consider other possibilities,
in particular, the case in which the neighbour is controlled by the
exponential decay with the $e^{-r}$ profile. In this situation, since
the exponential decay is much stronger than in the $1/r$ case, the
corresponding tail does not leave room for interaction with the
neighbour if it belongs to the next to next nearest neighbourhood. For
this reason, in this case, we have just one new model, the $c_2$
model, in which the mobility is implemented in the presence of two
distinct neighbourhoods, the first one with the nearest neighbours,
controlled by the probability $0.73$ and the second one, with the next
nearest neighbours controlled by the probability $0.27$. These
probabilities are taken to make the model follow the $e^{-r}$ profile.
The novelty here is then the presence of a model in which the first
and second neighbours follow the exponential decay; since it is
different from the other cases, in Fig. \ref{fig8} we display the
corresponding characteristic length, together with cases of the $a_1$
and $a_2$ models already studied, for comparison. The results are
obtained for $m=0.5$ and $N=1000$ and they show that the
characteristic length of model $c_2$ differs quantitatively, but it is
qualitatively similar to the other models. {\color{black} The qualitative agreement between the $1/r$ behavior and the $e^{-r}$ decay suggests that we search for a general rule for the mobility, which incorporate the two possibilities altogether. This is part of a new investigation, to be reported elsewhere.}

\section{Asymmetrical Mobility}
\label{other}

\begin{figure}[!htb]
	\centering
	\includegraphics{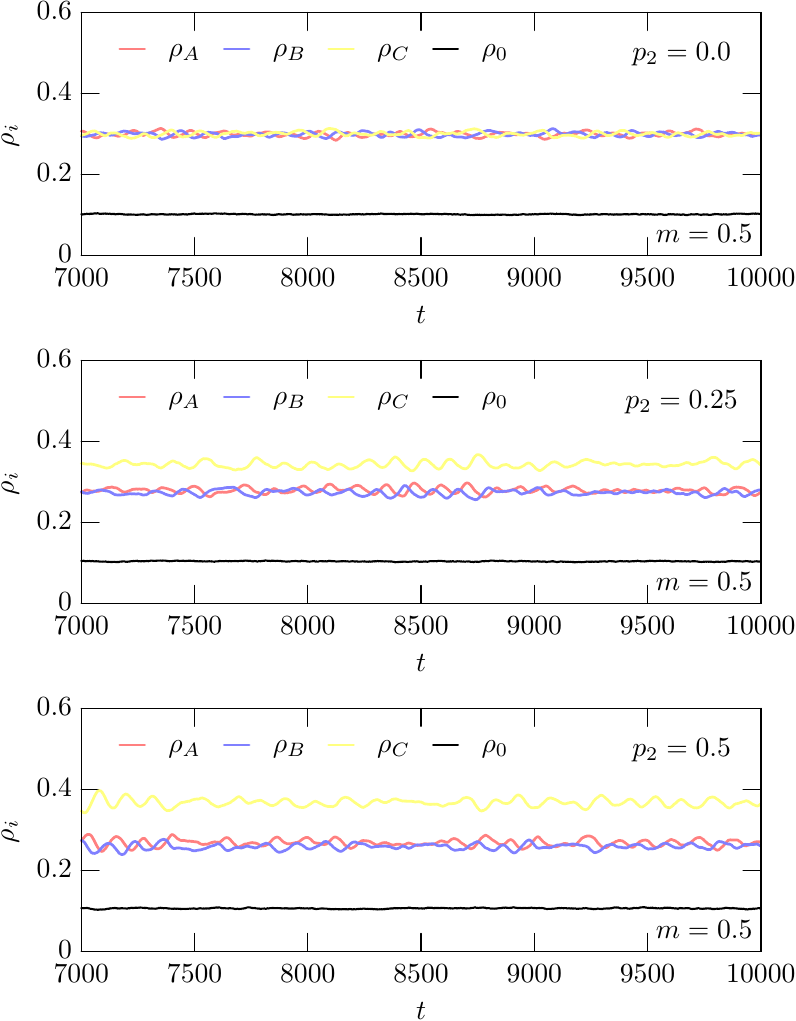}
	\caption{The abundance or density of the three species and the empty sites as a function of the generation, for three distinct values of $p_2=0, 0.25$ and $0.5$, in a lattice with linear size $N=1000$.}
	\label{fig9}
\end{figure}

The above investigations show that the neighborhood associated to the rule of mobility modify the characteristic length of the system, and may contribute to jeopardize biodiversity. Motivated by this fact and by the recent study \cite{2020-Bazeia-CSF-141-110356} where one breaks the unidirectional invasion of the species to jeopardize biodiversity, let us now consider other possibilities to change the standard rules of cyclic non hierarchical evolution. To implement new studies of current interest, let us now focus on having just one of the three species with a different mobility behavior. Among several possible modifications, we illustrate this idea choosing that species $A$ (red) can move to one of the 4 nearest neighbours with probability $p_1$, and to one of the 8 next nearest neighbours with probability $p_2=1-p_1$. The other two species keep the standard behavior, that is, they can only move to one of the 4 nearest neighbours with the same probability. Evidently, this creates an asymmetry in the cyclic environment of the model, and may be studied following the lines of Ref. \cite{2020-Bazeia-CSF-141-110356}. To see how this works in the present context, we depict in Fig. \ref{fig9} the abundance of the three species and the empty sites in a long run for some distinct values of the probability $p_2$. We choose, in particular, the cases $p_2=0$,  $0.25$ and $0.50$. The results show that the increase in the asymmetry favors species $C$ (yellow), increasing its abundance.  As we can see from Fig. \ref{fig9}, the raise of $p_2$ increases the power of the $A$ species to outcompete the $B$ population, diminishing its abundance or density. Consequently, the power of the $B$ species to outcompete the $C$  individuals is impaired, therefore increasing the abundance or density of the $C$ population. Due to the reduction of the $B$ population, the $C$ individuals increases in the asymmetric environment; in addition, the increased diffusion of the $A$ individuals makes it more vulnerable, allowing the $C$ predation to occur more effectively, negatively affecting the $A$ population. The enlargement of the abundance or density of the $C$  individuals that appears in Fig. \ref{fig9} is compensated by the lowering in the abundance of both $A$ and $B$ populations.          

\begin{figure}[!htb]
	\centering
	\includegraphics{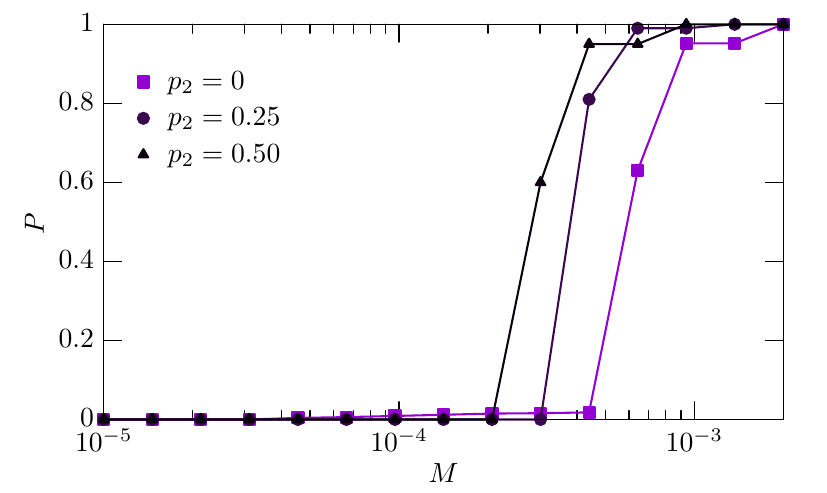}
	\caption{The extinction probability as a function of $M$, for the three values of the probability $p_2=0, 0.25$ and $0.5$, in a lattice with linear size $N=300$.}
	\label{fig10}
\end{figure}

This behavior seems to contribute to ease the extinction of biodiversity when one increases mobility to higher and higher values. For this reason, we further study the extinction probability in this asymmetric model and in Fig. \ref{fig10} we depict the results for  for the three distinct values of $p_2=0$, $0.25$ and $0.50$, with the results confirming the expected behavior, showing that the higher the value of $p_2$, the lower the value of $M$. This investigation is similar to the one displayed in the previous Fig. \ref{figX}, and here we have also checked that the linear size $N=300$ gives reliable results.

It is possible to explore the importance of the neighborhood in
another context, for instance, instead of enlarging the neighborhood
to account for the second and third neighbourhood, we can keep just
the first neighborhood, as usual, but change the way the species move,
reproduce or compete. In particular, we can use the rules of the model
$a_1$, and suppose that one of the three species, the $A$ species, for
instance, reproduces differently. If we do this for all the three
species, the overall effect is also of interest, but here we decided
to focus on breaking the cyclic symmetry present in the
rock-paper-scissors rules, so we suppose that only the individuals of
the species $A$ reproduce {\it considering all its four neighbors
simultaneously}. This means that under reproduction, when a species
$A$ site is selected, one immediately paints with the color red all
the empty sites at left, right, up and down positions. We can also
consider the case where the active site can simultaneously compete
with all adjacent sites; that is, under the competition rule the
active predator eliminates all preys eventually inhabiting its
neighborhood. We can consider the case of modified reproduction and
the case of modified competition separately or simultaneously. In the
three cases, we are breaking the cyclic symmetry present in the
standard model, and this will interfere importantly in the evolution
of the abundance of the species. {\color{black}These modifications are motivated by the recent study \cite{2020-Bazeia-CSF-141-110356} where one breaks the unidirectional invasion of the species, and also by similar investigations, recently implemented in a spatial model of bacterial competition \cite{pone}, with the identification of a range of values of growth rates of resistant cells where the species coexist, and the production mechanism is maintained as a polymorphism in the producing species, with the resistance mechanism also maintained in a similar way, as a polymorphism in the sensitive species. Another issue concerns the recent experimental results in simple microbial community with three strains of E. coli that interact cyclically \cite{2020-Liao-N-11-6055}, which describes in a specific E. coli community the presence of the survival of the weakest, as originally presented in Ref. \cite{2001-Frean-PRSLB-268-1323}}.

\begin{figure}[!htb]
	\centering
	\includegraphics{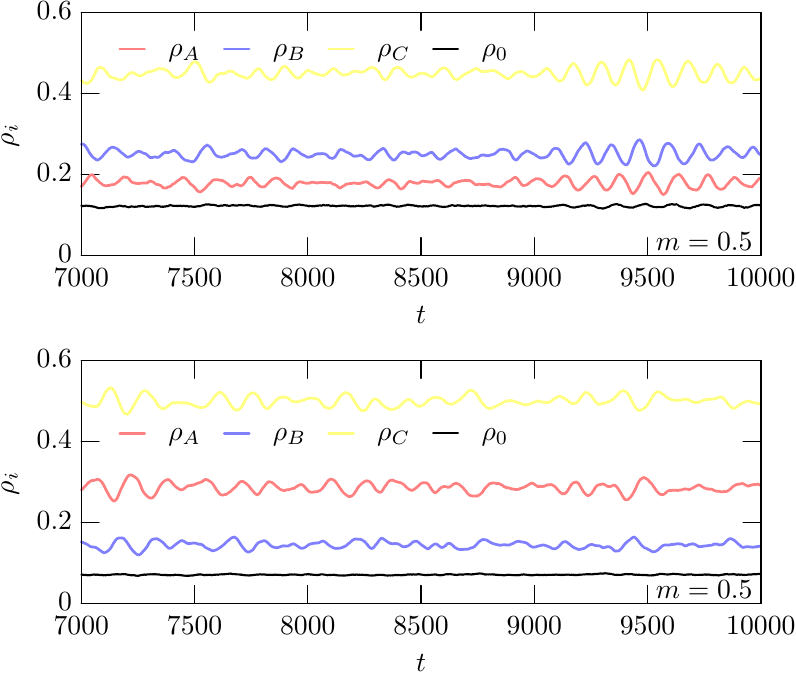}
	\caption{The abundance or density of the three species and the empty sites as a function of the generation, in a lattice with linear size $N=1000$. The top panel is for modified competition of the $A$  species, and the bottom panel is for modified reproduction of the $A$ species.}
	\label{fig11}
\end{figure}

We have studied these three distinct possibilities and in Fig. \ref{fig11} we depicted the abundance or density of the species and empty sites in the case with modified competition alone (top panel) and in the case of modified reproduction alone (bottom panel). The case where modified competition and reproduction act simultaneously leads to extinction of biodiversity, so we have omitted to display this behavior. In Fig. \ref{fig11} we see that when one of the two modified rules is at work, the $C$ population is the most abundant. This means that if one increases the power of one of the three species, the system ends up increasing the number of individuals of its competing species. In this sense, the result is similar to the previous case, displayed in Fig. \ref{fig9}. Here, however, the $A$ and $B$ species have distinct abundances and when the modification is in the competition, the $A$ species becomes itself the less abundant one. This is interesting, and shows that when one increases the power of competition of one of the three species, its abundance or density decreases the most. In this sense, we may say that the strongest individuals have the less abundant population. Although the result is different, it remind us of the survival of the weakest \cite{2001-Frean-PRSLB-268-1323}, which shows that the weakest individuals may constitute the most abundant population; see also the recent work \cite{2020-Liao-N-11-6055} for some experimental results on the survival of the weakest.

\section{Conclusion}
\label{conclusion}

In this work we considered a model containing three distinct species that evolve under the influence of the three standard rules known as mobility, reproduction and competition, with competition being implemented in accordance to the cyclic and non hierarchical rules of the rock-paper-scissors game. The competition between predator and prey is regulated by the May-Leonard rule, with the prey being eliminated leaving its corresponding site empty. In this simple and well-known model, the rules of mobility, reproduction and competition are all implemented under the guidance of the von Neumann neighborhood, only considering the first neighbor to provide interactions among all the individuals in the square lattice that obeys periodic boundary conditions. This model is revisited and used for comparison, as a reference for the modifications that we implemented in the other models, where mobility is changed to include the second and the second and third neighbours.

As we have shown, the increase of the neighborhood contributes to increase the characteristic length of the model. This was studied and quantified in this work. We have also shown that the characteristic length increases as we increase the value of the mobility. This behavior is quantitatively different but qualitatively similar for all the five models investigated in the work. The main result here leads to the interpretation that the larger the neighbourhood, the greater the characteristic length for each family of models.

Since the characteristic length increases to larger and larger values as we increase the mobility, we have also studied the extinction probability, that is, the vanishing of biodiversity as one increases mobility towards unity. The results show that the value of the extinction probability increases as we increase the neighborhood associated to the rule of mobility. This means that the increase in the motion of the individuals that compose the three species contributes to jeopardize biodiversity, adding to the related literature as a novel result of current interest. Moreover, motivated by the recent work \cite{2020-Bazeia-CSF-141-110356}, we have also investigated the possibility of breaking the cyclic symmetry of the model. Since the neighborhood is important to control biodiversity, we included an asymmetry in the model, considering one of the three species to have a larger neighborhood, when compared to the other two species. As we have shown, this possibility breaks the cyclic symmetry and favors one species to be the most abundant, also contributing to jeopardize biodiversity. Interestingly, the increase in the neighbourhood of one of the three species does not favor this species, but it in fact contributes to increase the abundance of its predator. It seems that the enlargement of the neighbourhood of one of the three species increases its associated predation, and this ends up increasing the abundance of its predator, within the cyclic environment governed by the rules of the rock-paper-scissors game.

Other results of current interest concern modifications in the manner the species compete and reproduce. These possibilities were considered with the breaking of the cyclic symmetry of the rock-paper-scissors game, and they ended up describing time evolutions with distinct abundances. As a particularly interesting result, it is sometimes possible to increase the power of competition of one of the three species, contributing to decrease its population the most. This means that the strongest species may sometimes generate the less abundant population.

The new results concerning the use of mobility to control biodiversity
maintenance show that mobility is a very efficient parameter to
investigate the time evolution of the numerical simulations which are
in general considered in the simple non hierarchical models based on
the rock-paper-scissors game. In this sense, the present investigation
suggests new studies, some of them related to more general models
where the three species are extended to describe four, five and other
quantities of species. A particular possibility refers to the
rock-paper-scissor-lizard-spock model, which is a generalisation the
rock-paper-scissor model to the case of five species. {\color{black}The modification of the rule of mobility to make it more general, including new possibilities such as the presence of Levy flights, and the change in the rules of reproduction and competition, to allow them to simultaneously interact with several neighbours are also of current interest.} We are now studying some possibilities, hoping to report on them in the near future.

\section*{ACKNOWLEDGMENTS}

Work partially financed by Conselho Nacional de Desenvolvimento Cient\'ifico e Tecnol\'ogico (CNPq, Grants nos. 303469/2019-6 and 404913/2018-0, Funda\c c\~ao Arauc\'aria, INCT-FCx (CNPq/FAPESP),  and Para\'iba State Research Foundation (FAPESQ-PB, Grant no. 0015/2019).

\bibliographystyle{elsarticle-num-names}

\end{document}